# A trust-based security mechanism for nomadic users in pervasive systems

Mohammed Nadir Djedid [1]

[1] Department of computer science, University of Sciences and Technology of Oran
-Mohamed Boudiaf-
Oran, 31000, Algeria

**Abstract**
The emergence of network technologies and the appearance of new varied applications in terms of services and resources, has created new security problems for which existing solutions and mechanisms are inadequate, especially problems of identification and authentication. In a highly distributed and pervasive system, a uniform and centralized security management is not an option. It then becomes necessary to give more autonomy to security systems by providing them with mechanisms that allows a dynamic and flexible cooperation and collaboration between the actors in the system.

## 1. Introduction

The rapid development of mobile computing has given rise to ubiquitous information systems: the user has at any time, access to the global network regardless of location or time ("anywhere - anytime" access).
"Pervasive" is a term that can be approximated with the idea of ubiquity or flooding. The idea of pervasive network would evoke a ubiquitous network.
The Ubiquitous (or pervasive) network allows a permanent connection of communicating devices that automatically recognize and locate themselves together, because they are "intelligent" objects. It evokes the notion of ambient intelligence.
The objects of the system are thus able to identify, store and interact naturally. But this use is made possible thanks to the wired and wireless connections, and also thanks to the possibility of interaction of objects together.
The challenge of pervasive systems is in this perspective, to provide methodological frameworks and protocols to permit the reliable, relevant and efficient use of these systems [1].
The large diversity of means of connection and communication conditions, added to that, the heterogeneity of devices, displays the data to the attacks on several levels. This requires, in fact, to provide tools for securing the system and to protect the identity of users. But this new trend reveals new security problems for which solutions and existing mechanisms are inadequate, especially for the problems of identification and authentication. In a highly distributed and pervasive system, a centralized and homogenous security management is not conceivable. It then becomes necessary to give more autonomy to security systems, providing them with mechanisms allowing a dynamic and flexible cooperation and collaboration between the actors in the system. This paper will be an overview of the main existing security systems and compare their effectiveness and their ability to meet the constraints of our problem.

We will present a first step, the requirements in terms of constraints and security needs of pervasive environments. Then, we will describe methods for authorization and standard modules used in the design of a distributed and collaborative security architecture. Subsequently, different solutions proposed for identification and authentication management will be studied, and their ability to meet the needs shown in the third part of the evaluated paper. Finally, a generic architecture of a security mechanism based on reputation and trust level will be proposed.

## 2. Security needs of pervasive systems

2.1 Decentralization

Security policy in a distributed system must be as decentralized as possible. Indeed, the user must be able to prove his/her identity anywhere in the environment without seeking systematically the centralized server of its parent organization.

2.2 Interoperability and Interaction

With the expansion of computer networks, the interaction, the federation and the cooperation between different organizations are prime prospects. The heterogeneity is one of the basic characteristics of pervasive applications. To design a security system allowing interaction and collaboration between disparate organizations is the closest alternative to reality.

## 2.3 Trust spread

The notion of trust is widely used in existing security systems, which allows delegating the mechanisms management of identification, authentication and authorization to several terminals. For example, a user is acquiring rights of access to organisms resources other than his/her parent organisms as long as these organizations know and trust each other, they can then work safely. But if that user moves to another organism, which is not known of his/her parent organism; he/she may be denied access to the resources, despite his/her status.

## 2.4 Traceability and non-repudiation

With the expansion of computer networks, the interaction, the federation and the cooperation between different organizations are prime prospects. The heterogeneity is one of the basic characteristics of pervasive applications. To design a security system allowing interaction and collaboration between disparate organizations is the closest alternative to reality.

## 2.5 Autonomy

Each user should be able to move independently across pervasive domains and to acquire rights without going through his/her original organism.

## 2.6 Transparency and Proactivity

In a ubiquitous environment, each entity must be able to authenticate itself and to acquire rights in an easy and transparent way. In addition, the resources used must seek the user as seldom as possible.

## 2.7 Flexibility

Certificates are the most used to identify an entity in a distributed system. However, the emergence of new technologies such as biometrics, RFID, forces the pervasive systems to integrate different means of identification, and adapt authentication mechanisms in relation to the context of the user, such as the ability and the type of the device used.

## 2.8 Privacy protection

In pervasive environment, a large amount of information (public or private) flows in networks. These data can provide information on preferences and the user's behavior, and can be misused. To remedy this fault, the user's device must have the ability to recognize the environment in which they are located, and to evaluate its degree of confidence. Only necessary and sufficient information for the identification / authentication will be available.

## 2.9 Scaling

One of the main challenges in pervasive systems is to allow interaction between different organizational entities. In other words, a pervasive security system is able to scale as long as it can accept an increasing number of new users.

# 3. Standard security modules

This section will aim to define certain types of authorization modes and standards security modules, required for establishing a security architecture manages the identification, authentication and access control, in order to understand the functioning of the different security systems that will be presented later.

## 3.1 Authorization modes

Three modes of authorization have been proposed by RFC 2904 [2]: agent mode, Pull mode, and Push mode.

- Agent mode

The authorization server acts as an intermediary (agent) between the user and the resource. The request of the user and the response of the resource go through the authorization server, which decides whether to allow access or not to the resource according to the policy associated to the user.

- Pull mode

In this mode, the resource requested by the user is directly sought. This refers to its authorization server to check whether this user is allowed to use it. Thus, if the answer is positive, the resource provides access to the user depending on the rights approved by the authorization server.

- Push mode

In this model, the user acts as an intermediary. A contact is first established with the authorization server, which delivers to the authorized user an access token (certificate); then this certificate is used as proof of the user's rights to utilize the resource.

## 3.2 Generic modules of a security system

From the RFCs (2904, [2] 3084 [3]), we can decompose a Security System Into Four Main modules:

- PIP (Policy Information Point): This section contains all data relative to users, resources, environment, etc…

- PAP (Policy Access Point): The policy of access control is defined in this module; it describes the rules to assign rights to each entity of the domain.
- PDP (Policy Decision Point): This module is in charge of processing requests, it first performs the collection of certain information related to the request from the PIP and PAP, and subsequently decides to accept or reject this request.
- PEP (Policy Enforcement Point): Plays the role of an interface of the security policy. It is responsible for receiving the users' requests, manage authentication and return the response made by the PDP.

## 4. Identity and privilege management systems

This section will be a state of the art of the main security systems providing the authentication and the access control. Existing solutions can be categorized into two sub-sections: Systems based on identity management, and systems that manage the authentication mechanism and the privileges.

4.1 Systems based on identity management

4.1.1 Radius [4]

The user sends an Access-Request containing his/her authentication information, and sends it to the server. The server processes the request locally if it recognizes the user, otherwise, it acts as a RADIUS Proxy "or intermediate" by transmitting it to another server. Exchanges are made via the chain of Radius Proxy servers intermediate in one direction and then in the other. When the request arrives at the Radius server corresponding the identification item, it validates the request or refuse it by sending a Access-Accept or Access-Reject packet. RADIUS operates in agent mode. Permissions management exists but remains basic. The main advantage of RADIUS is the deployment of inter-domain relations, by propagating information from one server to another. Otherwise, the decisions remain binary (Accept, Reject) and trust between the intermediate servers is not taken into account.

4.1.2 LemonLDAP[5]

Two connection modes exist for this approach, the pull mode, and the agent mode. In the pull mode, when a user wants to access a protected application, the system asks the user's name and password. Thus, after a successful authentication, the user is redirected to the resource that he/she seeks. In agent mode, the authenticated user accesses a menu containing all the applications on which he has access permissions. LemonLDAP is suitable for small organizations. Its centralized Single Sign On management prevents scaling and its inability to ensure interoperability between different security policies makes this solution unsuitable for large structures.

4.1.3 OpenID

OpenID [6] is a solution that permits to federate unique authentications and share attributes. It provides the ability to authenticate to multiple sites using a unique identifier OpenID. This model works in pull mode, and required to establish beforehand, trust relationships between service providers (websites, forums, messaging using OpenID) and identity providers (OpenID providers)..
The main advantages of OpenID are its ability to protect the users' attributes and traceability of the acount movement that can benefit the owner.
Otherwise, OpenID requires to the new users to drop all their existing accounts by creating a new account managed by their new supplier. Finally, OpenID is not working on push mode (offline).

4.1.4 Liberty Alliance

Liberty Alliance [7] proposes to combine the requirements of strong authentication (authentication of multiple attributes) by respecting the user's privacy. Just like OpenID, Liberty Alliance allows the users with a single account to access multiple services from different providers, but under the condition that they must belong to the same "circle of trust". The strong point of Liberty Aliance, is the separation between identity providers and service providers, which gives them more freedom to administer their resources. However, Liberty Aliance accuses the same problems suffered by OpenID.

4.1.5 Shibboleth

Shibboleth [8] is an authentication mechanism, which adopts the pull mode in its allocation of access rights. It permits to federate the identification and supply, as the mechanisms presented above, two possible applications: authentication delegation and sharing attributes. Shibboleth provides a cooperative aspect that makes his great strength. Contrary to what was seen previously, a user in this system, can use his/her original accounts to authenticate in any organization that trusts its parent organization, without being obliged to create a new account. However, the collaboration is limited to parties that have established explicit trust relationships, fault of consideration of extending the security policy in the design phase.

### 4.1.6 WS Security

WS-Security [9] is a security protocol called "point to point", which is dedicated to the message exchange of information between web services. Based on a mechanism of security tokens, it is associated with digital signatures to authenticate messages. Security tokens provide the identity of the message sender, which is proved by an authentication mechanism. The goal of WS-Security is to provide the users with a transparent and flexible authentication protocol, allowing different authentication modes from different organizations to interact. In addition to the ability to interconnect different security policies. This means, a user is authorized to use a service just if there is a rule that gives him/her the right in a explicit manner. But, the service administrator cannot know all the entities of the environment, which restricts access to a subset of users known directly.

## 4.2 Systems based on privilege management

### 4.2.1 Akenti

Akenti [10] is an architecture designed to provide security services in a completely distributed environnement. It defines two types of certificates:
- X.509 identity certificates: to identify a user.
- The attribute certificates, which are divided into two sub-categories:
– User certificates: containing the conditions of access to a resource.
– Authorization certificates: that list for each resource the various administrators authorized to create certificates of use for the concerned resource.

This mechanism works in push mode, in which the user requests a certificate authority, granting him/her the right to use the resource. The strength of Akenti is the autonomy offered to the user who has the right to negotiates access to a resource, by using authorization certificates. However, its limitation lies in the fact that for each resource, the user will need a certificate if he wants to use the push mode. In addition, the spread of trust is not implemented by Akenti.

### 4.2.2 Permis

Permis [11] includes a mechanism of static authority delegation. Thus, each actor defines trust authorities having the right to assign roles. In addition, a new version of Permis (2006), can delegate authority dynamically, by creating a chain of delegation. The spread of trust by the chain of delegation is considered as a breakthrough in the Permis project. It allows the extension of security policies, but obliges the authorities delegated to describe manually the trusted entities who can take advantage of privileges by the delegation.

### 4.2.3 CAS

CAS (Community Authorization Service) [12] is a protocol dedicated to control management in virtual organizations (VO) like grid computing. CAS assumes the role of supreme authority of a virtual organization and allows to manage resources and users between organizations working together in a common project. The principle is to delegate the management of certain resources that belong to various member organizations of the same VO to CAS server which manages the virtual organization. A user affiliated with the VO requests a certificate, called "Proxy Certificate" awarded by the CAS server of the corresponding community. This allows any entity to verify the authenticity of the certificate and allows the holder to access the resource indicated by the same certificate. Cooperation between CAS servers is a real evolution and opens the door to the development of a collaborative security system. However, each VO has a centralized server that manages the correspondence (mapping) between a static member fields, and cannot create them or evaluate them dynamically.

### 4.2.4 Voms

Virtual Organization Management Service [13] closely resembles the CAS and can operate in both push and pull modes. The major difference lies in the authorization mechanism. Indeed, like if in CAS, the attributes concerning the list of roles and groups members of the VO are stored in the voms server, the authorization rules are presented in the resource, that obtains the power to decide the user's right. In summary, this system considers the resource owner (not the administrator of the , as in CAS) is the entity responsible for the implementation of his/her security policy. However, the implementation of the VO is done via a centralized server as in CAS, and the interconnection of the VOs is not supported by voms, weakening the scope of the interconnection mechanism.

### 4.2.5 O2O

O2O [14] (Organization to Organization) is a security system for building a VO from several VPOs (Virtual Private Organizations). Like a VPN (Virtual Private Network), an VPO creates a bridge between two organizations. The policy of access control uses the same federation mechanism as Liberty Aliance, so that a unique profile can be attributed to each member an organization (the profile assigned by the original organism) and can thereby take advantages of the privileges with the organizations linked to this VPO gateway. As in the CAS system, the formation of a VO is achieved by gathering the different policies of each VPO in a centralized server to avoid conflicts that can cause alliances. O2O is an interesting mechanism in the way that it allows to build links between organizations in a decentralized manner. However, the extension of the system by transitivity links between VOs, through a centralization which is restrictive for the collaboration between them.

### 4.2.6 Sygn

In sygn [15] permissions are defined in the form of certificates stored at the owner. For the creation of such permissions, no interaction with a centralized system is necessary, which makes it one of the Sygn's strengths. Sygn also offers the possibility to define a permission on a set of resources. Sygn set up a decentralized architecture and reduces the interaction with trusted third parties. However, and as Permis system, the extension of trust occurs only through the delegation.

### 4.2.7 GAIA OS

GAIA OS Security [16] authenticates the user through different devices and protocols. A number between 0-1 is assigned to each device after authentication, which represents a measure of trust in the device or protocol. Thus, for a user to improve his/her reputation in the eyes of the system, he/she can combine several protocols to confirm his/her identity in order to increase the trust value associated with his/her (strong authentication). The advantage of GAIA OS, and in contrast to the mechanisms seen previously, is that the entity is measured digitally and not binary.
In addition, the collaboration between organizations is not taken into account, the delegation is the only supported, which reduces the application of security policy to the local domain only.

### 4.2.8 TrustAC

Trust-Based Access Control system [17] can control access through a mechanism estimating trust. In this system, the device of the user maintains its own security policy. Each person keeps a list of entities trusted or suspicious, by giving them a note of confidence that reflects their reputation with other users. Each person is not assigned to a role, but a number from 0-1, reflecting the confidence level of the individual within the community.
Otherwise, the system cannot evaluate the identity of the original organization and does not take into account the context of the individual, for exemple, the device used (Smartphone, laptop, etc.).

## 5. Synthesis

This section will be dedicated to the comparison between the different approaches presented above, in relation to the needs identified in Section 3. The result of the comparison is summarized in two tables. The following two mentions are used:
- Yes: The need is supported by the solution.
- No: The solution is not adapted for this need.

Table 1: Comparison of systems based on identity management

| Constraints | Systems based on identity management. | | | | | |
|---|---|---|---|---|---|---|
| | RADIUS | Lemon | OPENID | Liberty | Shib | WS-Sec |
| Decentralization | Yes | No | Yes | Yes | Yes | Yes |
| Interoperability | Yes | No | Yes | Yes | Yes | Yes |
| Trust spread | Yes | No | No | No | No | Yes |
| Traceability | Yes | Yes | Yes | Yes | Yes | Yes |
| Autonomy | Yes | No | Yes | Yes | Yes | Yes |
| Transparency and proactivity | Yes | Yes | Yes | Yes | Yes | Yes |
| Flexibility | No | No | No | No | No | Yes |
| Privacy Protection | No | No | Yes | Yes | Yes | No |
| Scaling | Yes | No | Yes | No | No | Yes |

Table 2: Comparison of systems based on privilege management.

| Constraints | Systems based on privilege management. | | | | | | | |
|---|---|---|---|---|---|---|---|---|
| | Akenti | Permis | CAS | Voms | O2o | Sygn | GAIA. | Trust |
| *Decentraliza.* | Yes | No | Yes | No | Yes | Yes | Yes | Yes |
| *Interop.* | Yes | Yes | Yes | Yes | Yes | Yes | No | No |
| *Trust spread* | No | Yes | Yes | No | No | Yes | Yes | Yes |
| Traceab. | No | Yes | No | No | No | Yes | Yes | No |
| *Autono.* | Yes | Yes | Yes | Yes | Yes | Yes | No | Yes |
| *Transparency /proactivity* | No | No | No | No | No | No | Yes | No |
| *Flexibil.* | No | No | No | No | No | No | Yes | No |
| *Privacy Protec.* | No | No | No | No | No | No | No | No |
| *Scaling* | No | Yes | Yes | Yes | Yes | Yes | No | Yes |

## 6. A trust-based security mechanism for nomadic users

In order to fully exploit the concept of spreading the trust to interconnect security systems of various domains, we propose a generic architecture in which different modules are shown in Figure 1:

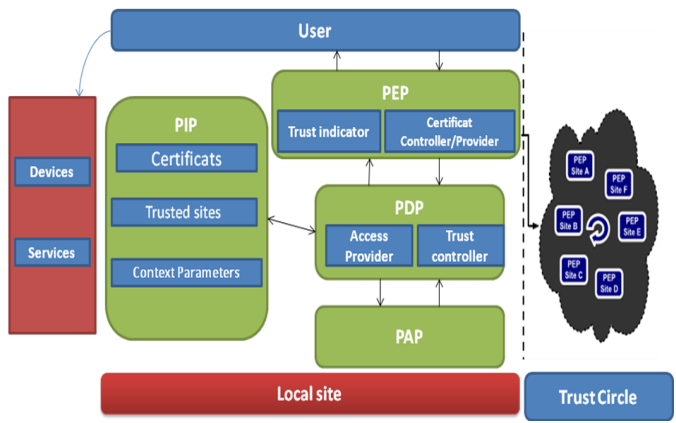

Fig. 1  A trust-based security mechanism for nomadic users

**PEP**: Is the external interface of the architecture through which pass all the information in the form of certificates, it has a particular module called "Trust indicator" and reflects the reliability level of the site, the site's reputation, the number of links of trust with other sites etc. All these information are used to assign a trust level to the site. The interface can also verify the veracity of the certificates exchanged between the system and the outside thanks to the controller of certificates, and provide certificates commanded by the PDP, to be sent to entities (user, device).

**PDP**: Allows filtering access to the system via the trust controller, and deciding to establish or revoke the trust with the sites. It also allows applying the security policy defined in the PAP module.

**PAP**: Is the module where the access control policy is defined.

**PIP**: Allows the capture of the user's context (device used, connection type etc.). It also maintains a table of trusted sites updated by the trust controller module of PDP.

## 7. Conclusion

Through this study we came to the conclusions that the concept of the propagation of trust in a dynamic way is not fully exploited to interconnect the security policies in various fields. Reply to this lack could bring us closer to our goal of protecting the identity of the users, and this in "globalizing" the SSO system across domains: a single sign-on (SSO) would not only provide access to several domain resources belonging to the user, but also the resources of the areas of trust where the user goes, without being forced to decline again its identity. This would avoid to re-circulate the information of identification / authentication at the risk that it would be intercepted by a third party.
Therefore, we proposed a generic architecture, setting up a trust-based security mechanism based on reputation and trust level accumulated by each domain towards its peers.
This work is a first step on designing our architecture, and the future works will be focused on calculating the value of the trust level by providing a function that calculates this value.